\documentclass{nic-series}
\begin{document} 
\title{Adsorption phenomena at organic-inorganic interfaces}
\author{Michael Bachmann \and Wolfhard Janke}
\institute{Institut f\"ur Theoretische Physik and Centre for Theoretical Sciences (NTZ), \\
        Universit\"at Leipzig, 04109 Leipzig, Germany\\
        \email{\{bachmann, janke\}@itp.uni-leipzig.de}}
\maketitle
\begin{abstracts}
The qualitative solvent- and temperature-dependent conformational behavior of a peptide
in the proximity of solid substrates with different adsorption properties is investigated
by means of a simple lattice model. The resulting pseudophase diagrams exhibit a complex
structure, which can be understood by analysing the minima of the free-energy landscape
in dependence of appropriate system parameters. 
\end{abstracts}
\section{Introduction}
The interest in understanding adsorption of polymers and peptides to inorganic material 
has grown recently with the development of nanotechnological equipment that enables
present and future applications ranging from nanoelectronic devices [as, for example,
the celebrated organic light-emitting diodes (OLEDs)] to pattern-recognising nanosensors
in biomedicine. In recent experiments, evidence for substrate specificity of synthetic peptide
adsorption to semiconductor substrates has been found.~\cite{bj2semi1} The investigation of 
organic-inorganic interfaces is connected with the substantial problem of setting up an effective 
microscopic model for the adsorption of organic to inorganic matter in addition to the intrinsic 
interactions governing the folding properties of the organic substance. Van der Waals based
interaction models~\cite{bj2hentschke1} were employed, e.g., in studies of gold-binding 
peptides.~\cite{bj2schulten1} Here, we study a minimalistic lattice model, where the folding
part is governed by the hydrophobic-polar (HP) model,~\cite{bj2dill1} and the interaction
with the substrate is restricted to monomer-type dependent surface-layer attraction.~\cite{bj2bj1,bj2bj1b}
\section{Conformational Transitions in a Simple Hybrid Interface Model}
Based on the assumption that peptide folding is mainly due to the hydrophobic effect, 
we distinguish only hydrophobic ($H$) and polar residues ($P$). According to the minimalistic
HP model~\cite{bj2dill1}, the intramolecular energy of a conformation is related to the number
of hydrophobic nearest-neighbor contacts $n_{\rm HH}$.
For the study of  specificity of surface--binding, we investigate three attractive substrate 
models: (a) type-independent attractive, (b) hydrophobic, and (c) polar substrate. 
In case (a), the energy of the system is proportional
to the total number of monomer--surface contacts, $n_s^{\rm H+P}$, while in the cases (b) and (c)
the respective hydrophobic ($n_s^{\rm H}$)
and polar ($n_s^{\rm P}$) surface contacts are energetically favored. The generic model
can be defined by $E(n_s,n_{\rm HH})=-n_s-s n_{\rm HH}$,
where, depending on the substrate model, $n_s=n_s^{\rm H+P}$, $n_s^{\rm H}$, or 
$n_s^{\rm P}$. The parameter $s$ controls the solvent quality (the larger
the value of $s$, the worse the solvent). The contact numbers $n_s$ and $n_{\rm HH}$
are natural system parameters that allow the discrimination of conformational pseudophases.
In our investigations of this system, we employed
the contact-density chain-growth method which has 
successfully been applied in adsorption studies of polymers.~\cite{bj2bj3} This method, 
which is a suitably generalized variant of the multicanonical chain-growth algorithm,~\cite{bj2bj2} 
allows us to estimate the contact density $g(n_s,n_{\rm HH})$, which simply
counts the number of conformations for a given pair of contact 
numbers $n_s$ and $n_{\rm HH}$. For fixed temperature $T$ and solubility $s$,
we define the contact free energy $F_{T,s}(n_s,n_{\rm HH}) \sim -T\ln\, g(n_s,n_{\rm HH})\exp(-E_s/T)$.
Assuming that the minimum of the free-energy landscape $F_{T,s}(n_s^{(0)},n_{\rm HH}^{(0)})$ 
is related to the class of macrostates with $n_s^{(0)}$ surface and 
$n_{\rm HH}^{(0)}$ hydrophobic contacts, this class dominates the pseudophase the system resides in. For this
reason, it is instructive to calculate all minima of the contact free energy and to determine the 
associated contact numbers in a wide range of values for the external parameters. This can easily be done 
with the knowledge of $g(n_s,n_{\rm HH})$ by simple reweighting. 

\begin{figure}
\centerline{ \epsfxsize=8.2cm \epsfbox{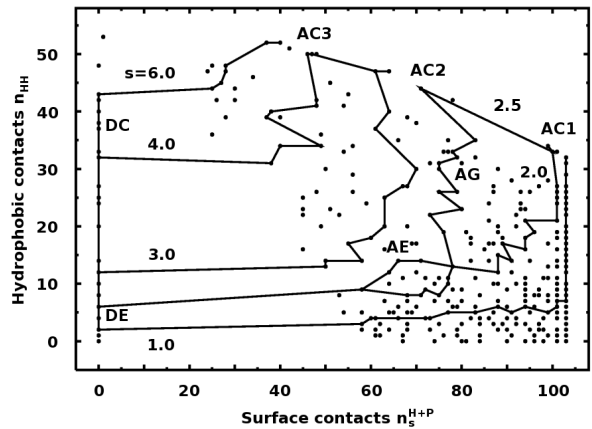} }
\caption{\label{fig:bj2lm}  Contact-number map of all free-energy minima for an exemplified peptide
with 103 monomers~\cite{bj2bj1}
and a substrate which is equally attractive to all monomers. 
Circles correspond to minima of the contact free energy
$F_{T,s}(n_s^{\rm H+P},n_{\rm HH})$
in the parameter space $T\in [0,10]$, $s\in[-2,10]$. Lines illustrate how the contact free energy changes
with the temperature at constant values of the solvent parameter $s$. }
\end{figure}
In Fig.~\ref{fig:bj2lm}, the map of all possible free-energy 
minima (circles) in the range of external parameters $T\in[0,10]$ and $s\in[-2,10]$ is shown for
a peptide with 103 monomers (37 hydrophobic, 66 polar) in the vicinity of a substrate that 
is equally attractive for both hydrophobic and polar monomers.~\cite{bj2bj1,bj2bj1b}
There are a few regions of circles which can be associated with pseudophases. Clearly, where the number of 
surface contacts $n_s^{\rm H+P}$ is zero, the heteropolymer is desorbed and in the bulk regime
two main pseudophases can be identified: desorbed extended (DE) and desorbed collapsed (DC) hydrophobic-core  
conformations. In case of adsorption, we distinguish also extended (AE) and compact (AC1-3)
conformations, and an intermediate globular pseudophase (AG), where conformations are globally
compact, but do not exhibit pronounced hydrophobic domains. The discrimination of the three AC 
subphases is due to layering: In dependence of the solvent quality, hydrophobic domains form
layers: single-layer domains in AC1 (good solvent, substrate contact favored prior formation of
intrinsic hydrophobic contacts), and respective hydrophobic double- and triple-layer cores
in AC2 and AC3.  
Solid lines in Fig.~\ref{fig:bj2lm} visualize ``paths'' through the free-energy landscape when 
changing temperature, but leaving the solvent parameter unchanged ($s={\rm const}$). Recalling
that larger $s$ values correspond to worse quality of the solvent, it is obvious that the $s=6$ path
at low temperatures starts in the highly compact AC3 subphase. Increasing the temperature, 
the monomers reorder and the system enters for entropic reasons the globular pseudophase AG. 
Approaching $T\approx 2.3$ a noticeable first-order-like unbinding transition occurs towards pseudophase
DC, before the compact structure decays (DE). Under very good solvent conditions as, e.g., for the
trajectory of $s=1$, the favored conformations at low temperatures are film-like. Increasing 
the temperature slightly to $T\approx 0.2$, the system crosses over to the AE pseudophase. 
This subphase is very stable under these solvent conditions,
but close to $T\approx 2.3$ the heteropolymer desorbs into the dissolved bulk phase DE.
  
For the other two substrates, the pseudophase diagrams look quantitatively 
different but most of the main pseudophases are also present in these cases.~\cite{bj2bj1}
\section{Concluding Remarks}
Beside the expected adsorbed and desorbed phases, hybrid organic-inorganic systems exhibit a rich 
subphase structure in the adsorbed phases with compact hydrophobic domains. In this region, the binding
behavior is strongly substrate-specific and depends in detail on the quality of the solvent.
Since current experimental equipment can reveal molecular structures at
the nanometer scale, it should be possible to investigate the grafted structures
in dependence of the solvent quality.
Such experiments would be essential for studies of 
binding forces that are strong enough to refold peptides or proteins.
\section*{Acknowledgments}
This work is partially supported by the DFG (German Science Foundation) under Grant
No.\ JA 483/24-1 and the computer time Grant No.\ hlz11 of the John von Neumann Institute for 
Computing (NIC), Forschungszentrum J\"ulich.

\end{document}